\documentclass{svjour3}                     
\smartqed  
\usepackage{graphicx}
\journalname{Few-Body Systems (FB20)}
\begin{document}

\title{
Mechanism of {\boldmath $K^- p\to \eta\Lambda$} reaction near threshold
\thanks{Presented at the 20th International IUPAP Conference on Few-Body Problems in Physics, 20 - 25 August, 2012, Fukuoka, Japan}
}


\author{Bo-Chao Liu       \and
        Ju-Jun Xie 
}


\institute{B. C. Liu \at
              Department of Applied Physics, Xi'an
Jiaotong University, Xi'an, Shanxi 450001, China. \\
              \email{liubc@xjtu.edu.cn}           
           \and
          J. J. Xie \at
              Department of Physics, Zhengzhou University, Zhengzhou,
Henan 450001, China.
}

\date{Received: date / Accepted: date}

\maketitle

\begin{abstract}
Within an effective Lagrangian approach, we investigate the reaction
mechanism of $K^- p\to \eta\Lambda$ near threshold. It is
found that a new $D_{03}$ resonance, with mass $M=1668.5\pm 0.5$MeV
and width $\Gamma=1.5\pm0.5$MeV, is needed to interpret the
experimental data reported by the Crystal Ball collaboration. To verify
our results, remeasurements on the differential cross sections and
$\Lambda$ polarization in this reaction will be helpful.
Furthermore, we also show that the reaction $p\bar p\to \Lambda\bar
\Lambda \eta$ is a good place to look for this new resonance if it
exists.
\end{abstract}

\section{Introduction}
\label{intro} The $K^-p$ reactions are important methods for
studying the properties of hyperon resonances. Among of them, the
reaction $K^- p\to \eta\Lambda$ is particularly interesting, because
both $\eta$ and $\Lambda$ are isospin singlet, there is no
contributions from the $\Sigma$ resonances, and therefore, this reaction
is especially suitable for studying the properties of $\Lambda$
resonances. Even though the $\eta\Lambda$ channel is a relatively
clean channel for studying $\Lambda$ resonances, current knowledge
on the couplings of $\Lambda$ resonances with the $\eta\Lambda$ channel
is still not satisfying. In Particle Data Group(PDG)
book~\cite{pdg}, there is only one state, $\Lambda(1670)$, which has
well established decay branch to the $\eta\Lambda$ channel. The
couplings of other $\Lambda$ resonances with the $\eta\Lambda$ channel
are only very poorly known. This is possibly due to the weak
couplings of other $\Lambda$ resonances with the $\eta \Lambda$ channel.
Another possible reason is the limited early experimental data with
very poor quality.

In 2001, some new experimental data with much better precision than
before were reported by the Crystal Ball collaboration\cite{data}. These
new data offer a better basis for investigating the reaction
mechanism of $K^- p\to \eta\Lambda$ and extracting the information
about $\Lambda$ resonances in this reaction.

\section{Results and Discussions}
\label{sec:1}
\begin{figure*}
  \includegraphics[scale=0.4]{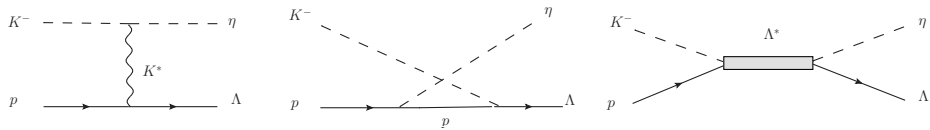}
\caption{Feynman diagrams of our model.}
\label{feynfig}       
\end{figure*}
 Within an effective Lagrangian approach, we
investigate the mechanism of $K^- p\to \eta\Lambda$ reaction near
threshold~\cite{prc,prcnew}. We consider $t-$channel $K^*$ exchange
and $u-$channel proton exchange as background contributions(as shown
in Fig.~\ref{feynfig}(a)). In $s-$channel, firstly we only include
$\Lambda(1670)$ exchange, because it is the only $\Lambda^*$ state
which has significant coupling to the $\eta\Lambda$ channel. We set the
coupling constants appearing in the amplitudes and the mass and
width of $\Lambda(1670)$ as free parameters. We also include two
parameters as the relative phase factors among amplitudes. Then we
fit these parameters to the data. It is found that although the
total cross section data can be well described, the bowl structures
shown in angular distributions cannot be reproduced at all(See
Ref.~\cite{prc,prcnew} for more details). This result is not unexpected as we only include one $s-$wave resonance in this fit. It
seems we still need some higher partial wave contributions.

Next, we include a new $D_{03}$ resonance in the fit with setting the parameters of
this resonance as free parameters. In this new fit, both total and
differential cross section data can be well described. The best
fitting results favor the $D_{03}$ resonance with mass
$M=1668.5\pm0.5$ MeV and width $\Gamma=1.5\pm0.5$ MeV, which is
obviously not a known $\Lambda$ resonance listed in PDG. Here it
will be interesting to discuss the role of $\Lambda(1690)$ in this
reaction. $\Lambda(1690)$ is a well established $D_{03}$ state and
also lies near $\eta\Lambda$ threshold. In fact, in the experimental
paper\cite{data} the authors also noticed the higher partial wave
contributions shown in the angular distributions, and they argued
that these higher partial wave contributions may come from
$\Lambda(1690)$. However, the largest beam momentum of these new
data is about 770 MeV, which corresponds to invariant mass
$\sqrt{s}=1.685$ GeV, the lower limit of the mass of $\Lambda(1690)$
suggested by PDG. Therefore if these higher partial wave contributions are
due to $\Lambda(1690)$, one may expect that the bowl structure shown
in the angular distributions should be more and more prominent with
increasing energies. Such an expectation is not supported by the data,
so we do not think these higher partial wave contributions are
caused by $\Lambda(1690)$.

\begin{figure}
\includegraphics[scale=0.2]{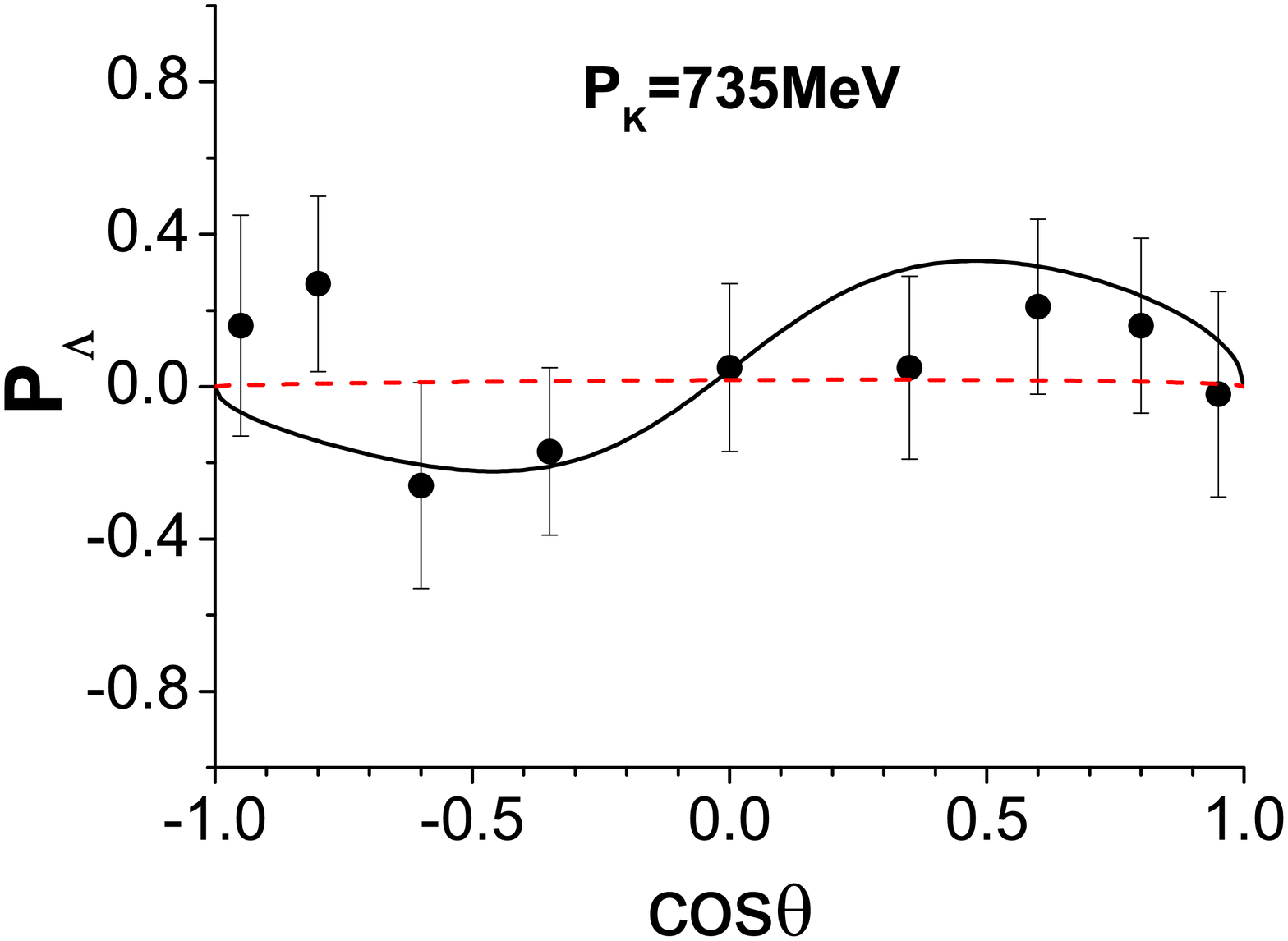}
\includegraphics[scale=0.2]{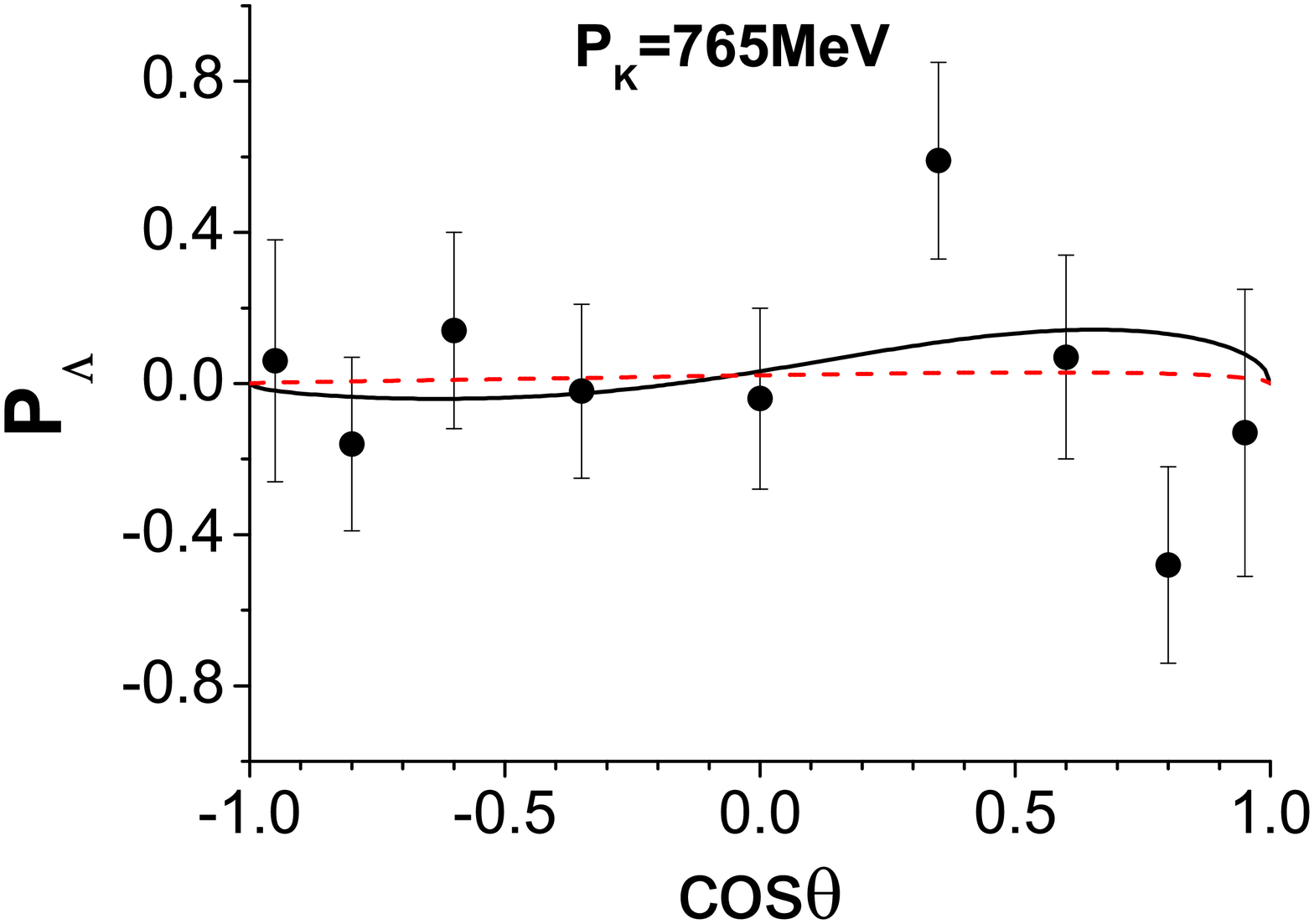}
\caption{The predictions for $\Lambda$ polarization with(black solid line) and
 without(red dashed line) including the narrow resonance. For details see Ref.\cite{prcnew}.}
\label{pol}       
\end{figure}

Because there are no other experimental evidences or theoretical
predictions for the existence of this narrow resonance, it will be important to find
some other proofs to establish the existence of this new resonance. Fortunately, there are some ways to verify our results. The
first way is from the $\Lambda$ polarization data. In fact, besides
the total and differential cross section data some $\Lambda$
polarization data are also reported by the Crystal Ball
collaboration in the same paper~\cite{data}. Because the quality of
the $\Lambda$ polarization data is poor and the number of these
data points is small, we did not include these data in the fit.
However, after fixing the free parameters by fitting to the total
and differential cross section data, it will be interesting and important to check
our model predictions for the $\Lambda$ polarization. The predictions of our model
for the $\Lambda$ polarization are shown in Fig.~\ref{pol}.
As can be seen from the figure, with large uncertainties of current
data one cannot get any decisive conclusions. However, it is shown
that with or without the narrow $D_{03}$ resonance the models give
distinct predictions for the $\Lambda$ polarization at
$P_{K^-}=735$MeV. Thus if we have more accurate $\Lambda$
polarization data in the future, we will have more evidence for
this narrow resonance.

Another way is the measurements on the $p\bar p\to
\Lambda\bar\Lambda \eta$ reaction. With the parameters fixed in the
$K^- p\to \eta\Lambda$ reaction, we can make some predictions for
the observables of $p\bar p\to \Lambda\bar\Lambda \eta$ reaction
based on meson exchange model(see Fig.~\ref{feynfig}(b)).

\begin{figure}
\includegraphics[scale=0.3]{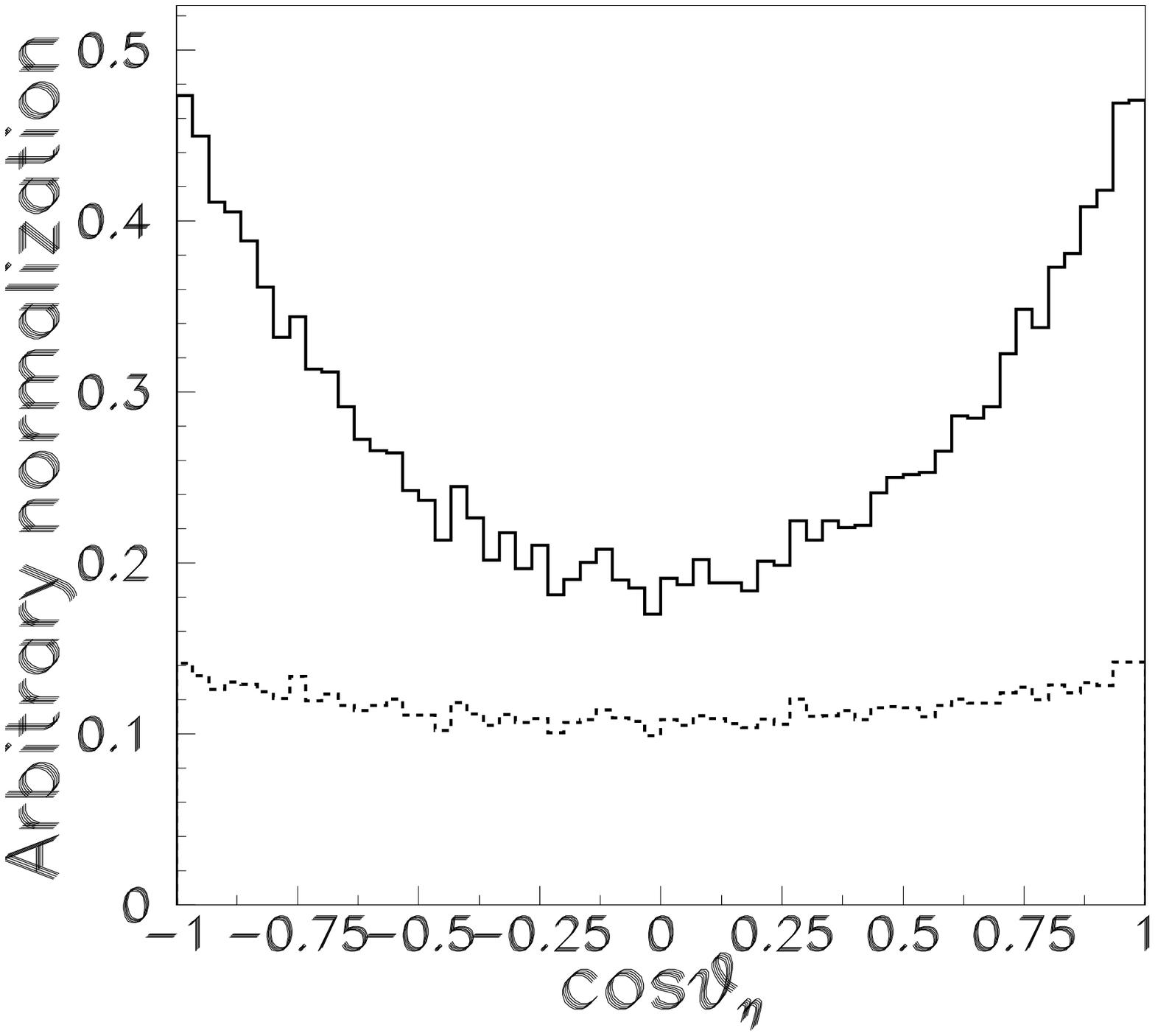}
\includegraphics[scale=0.3]{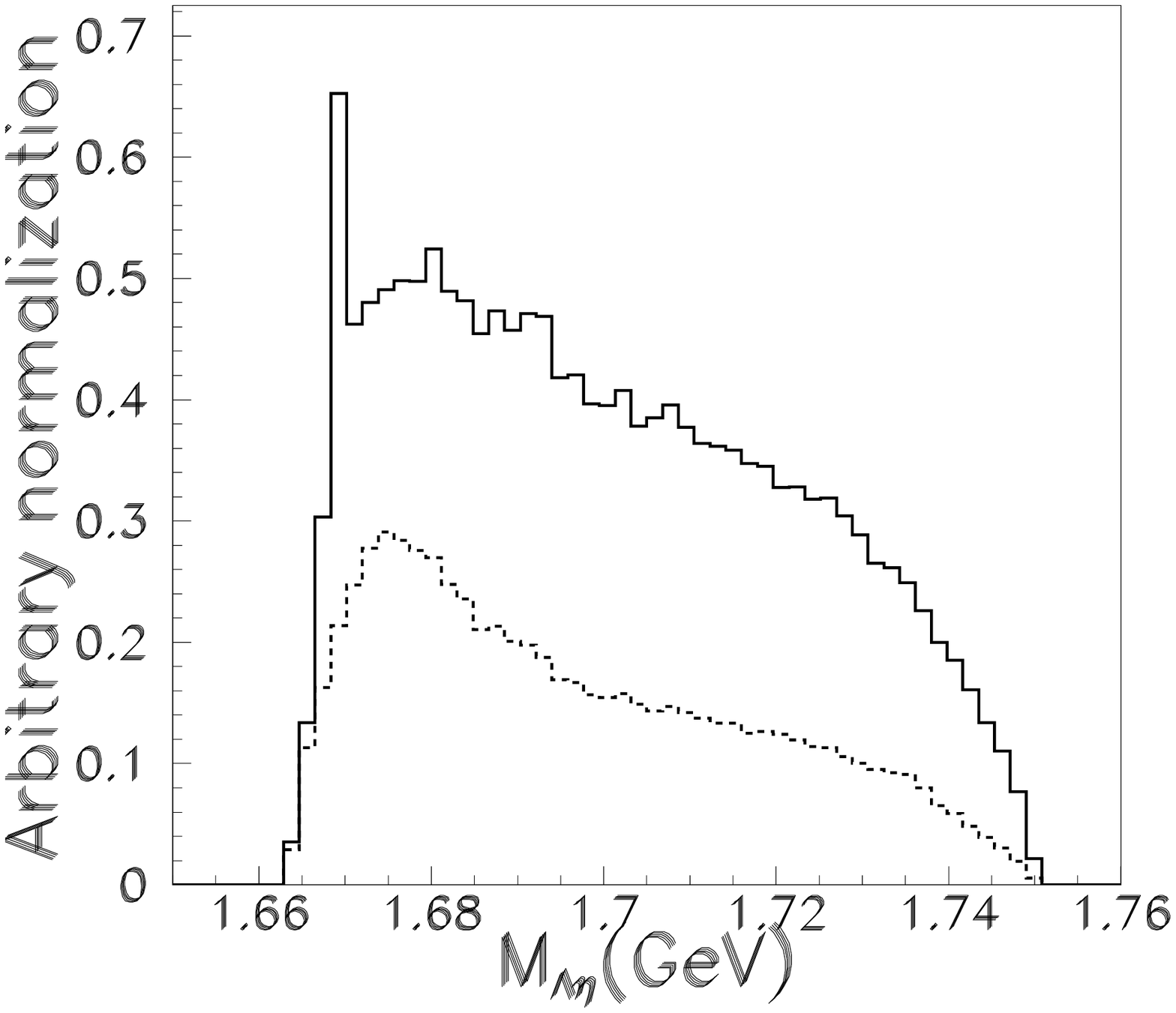}
\caption{Preliminary results for the predictions of the $p\bar p \to \Lambda\bar\Lambda\eta$ reaction with(solid line) and
 without(dotted line) including the narrow resonance.}
\label{ppbar}       
\end{figure}

In a simplified model, we ignore the initial and final state
interactions and we only consider kaon exchange. The calculated
results for the invariant mass spectrum of $\eta\Lambda$ and the
angular distribution of $\eta$ meson in center of mass frame are
shown in Fig.~\ref{ppbar}. There are mainly two findings from these
calculations. The first finding is that with or without the narrow
$D_{03}$ resonance the invariant mass spectrum of $\eta\Lambda$
system and the angular distribution of $\eta$ meson will show very
different pattern. The second finding is that, compared to the small
bump caused by the narrow resonance in the total cross sections of
the $K^- p\to \eta\Lambda$ reaction, the invariant mass spectrum of
$\eta\Lambda$ shows a sharp peak here. It indicates that the narrow
$D-$wave resonance's contribution is enhanced in this reaction. This
enhancement is due to some simple kinematical reason. Because the
threshold momentum is much larger for the $p\bar
p\to\Lambda\bar\Lambda \eta$ reaction than that for the $K^- p\to
\eta\Lambda$ reaction, the vertex function of $\Lambda^*\bar KN$ vertex
for $D-$wave resonance($\sim p^2$) will be largely enhanced in the
$p\bar p\to\Lambda\bar\Lambda \eta$ reaction compared to the $K^-
p\to \eta\Lambda$ reaction. For $S-$wave resonance, there is no such
enhancement. So the signal of the narrow $D-$wave resonance is
enhanced in the $p\bar p\to\Lambda\bar\Lambda\eta$ reaction, which
makes this reaction a good place to look for the narrow resonance if
it exists.

\section{Summary}
The main conclusions of this work can be summarized as
follows:
\begin{enumerate}
\item
With including the $\Lambda(1690)$, the higher partial wave contributions shown in angular distributions cannot be explained.

\item
With including a narrow $D_{03}$ resonance($M=1668.5\pm 0.5$MeV,
$\Gamma=1.5\pm0.5$MeV), we can describe the data well.

\item
Remeasurements on the differential cross section and $\Lambda$ polarization data and measurements on the reaction $p\bar p\to \Lambda\bar\Lambda\eta$ may help.

\item
The $\Lambda(1670)$ plays a dominant role near threshold, while the
background contributions are also important.

\end{enumerate}
We suggest experimentalists to perform the new measurements
and find further experimental evidences to establish the (non-)existence of this narrow
resonance.

\begin{acknowledgements}
We acknowledge the support of the National Natural Science
Foundation of China under grant no 10905046 and 11105126. B. C. Liu
is also supported by the Fundamental Research Funds for the Central
Universities.
\end{acknowledgements}



\end{document}